\documentclass[sigconf,nonacm]{acmart}
\AtBeginDocument{\settopmatter{printacmref=true}}

\usepackage{tabularx}
\usepackage{graphicx}
\usepackage{dblfloatfix} 
\usepackage{cuted}
\usepackage{capt-of}
\usepackage{cuted}
\usepackage[most]{tcolorbox} % for the colored box
\usepackage{xcolor}        
\usepackage{enumitem}        
\usepackage{pifont}         
\usepackage{cuted}
\definecolor{ideatitle}{RGB}{76,175,80}   % top bar
\definecolor{ideaback}{RGB}{232,245,233}  % body

\newtcolorbox{ideabox}[2][]{%
  breakable,
  enhanced,
  colframe=white,
  boxrule=0pt,
  left=2mm,right=2mm,top=0mm,bottom=2mm,
  arc=0pt,outer arc=0pt,
  colback=ideaback,
  colbacktitle=ideatitle,
  coltitle=white,
  fonttitle=\bfseries\itshape,
  fontupper=\itshape,
  title={#2},
  #1 % allow local overrides like colback=..., colbacktitle=...
}

\AtBeginDocument{%
  }

\setcopyright{acmlicensed}
\copyrightyear{2026}
\acmYear{2026}
\acmDOI{XXXXXXX.XXXXXXX}

\acmConference[FAccT'2026]{Proceedings of ACM Conference on Fairness, Accountability, and Transparency}{June 25--29, 2026}{Montreal, Canada}

\acmISBN{978-1-4503-XXXX-X/2018/06}

\begin{document}

\title[Designing Safe and Accountable GenAI with Women Banned from Formal Education]%
{Designing Safe and Accountable GenAI as a Learning Companion with Women Banned from Formal Education}

\author{Hamayoon Behmanush}
\authornote{{\color{blue}\textbf{This work has been accepted at ACM Conference on Fairness, Accountability, and Transparency 2026 as a full paper. Please cite the peer-reviewed version.}}}
\email{behmanush@cs.uni-saarland.de}
\affiliation{%
  \institution{Saarland Informatics Campus, Saarland University}
  \city{Saarbrücken}
  \country{Germany}
}

\author{Freshta Akhtari}
\affiliation{%
  \institution{Computer Science Faculty, Parwan University}
  \city{Parwan}
  \country{Afghanistan}
}

\author{Ingmar Weber}
\affiliation{%
 \institution{Saarland Informatics Campus, Saarland University}
 \city{Saarbrücken}
 \country{Germany}
}
 
\author{Vikram Kamath Cannanure}
\affiliation{%
 \institution{Saarland Informatics Campus, Saarland University}
 \city{Saarbrücken}
 \country{Germany}
}

% \author{Hamayoon Behmanush}
% \email{behmanush@cs.uni-saarland.de}
% \affiliation{%
%   \institution{Saarland Informatics Campus, Saarland University}
%   \city{Saarbrücken}
%   \country{Germany}
% }

% \author{Freshta Akhtari}
% \affiliation{%
%   \institution{Computer Science Faculty, Parwan University}
%   \city{Parwan}
%   \country{Afghanistan}
% }

% \author{Ingmar Weber}
% \affiliation{%
%  \institution{Saarland Informatics Campus, Saarland University}
%  \city{Saarbrücken}
%  \country{Germany}}
 
% \author{Vikram Kamath Cannanure}
% \affiliation{%
%  \institution{Saarland Informatics Campus, Saarland University}
%  \city{Saarbrücken}
%  \country{Germany}}

\renewcommand{\shortauthors}{Behmanush et al.}

% \authornote{{\color{blue}\textbf{This work has been accepted at ACM Conference on Fairness, Accountability, and Transparency 2026 as a full paper. Please cite the peer-reviewed version.}}}

\begin{abstract}
In gender-restrictive and surveilled contexts, where access to formal education may be restricted for women, pursuing education involves serious safety and privacy risks. When women are excluded from schools and universities, they often turn to online self-learning and generative AI (GenAI) to pursue their educational and career aspirations. However, we know little about what safe and accountable GenAI support is required in the context of surveillance, household responsibilities, and the absence of learning communities. We present a remote participatory design study with 20 women in Afghanistan, informed by a recruitment survey (n = 140), examining how participants envision GenAI for learning and employability. Participants describe using GenAI less as an information source and more as an always-available peer, mentor, and source of career guidance that helps compensate for the absence of learning communities. At the same time, they emphasize that this companionship is constrained by privacy and surveillance risks, contextually unrealistic and culturally unsafe support, and direct-answer interactions that can undermine learning by creating an illusion of progress. Beyond eliciting requirements, envisioning the future with GenAI through participatory design was positively associated with significant increases in participants’ aspirations (p=.01), perceived agency (p=.01), and perceived avenues (p=.03). These outcomes show that accountable and safe GenAI is not only about harm reduction but can also actively enable women to imagine and pursue viable learning and employment futures. Building on this, we translate participants’ proposals into accountability-focused design directions that center on safety-first interaction and user control, context-grounded support under constrained resources, and offer pedagogically aligned assistance that supports genuine learning rather than quick answers.

\end{abstract}




%   \includegraphics[width=\textwidth]{Figures/TeFig.png}
%   \caption{Expa....}
%   \Description{...}


\begin{teaserfigure}
  \centering
  \includegraphics[width=\textwidth]{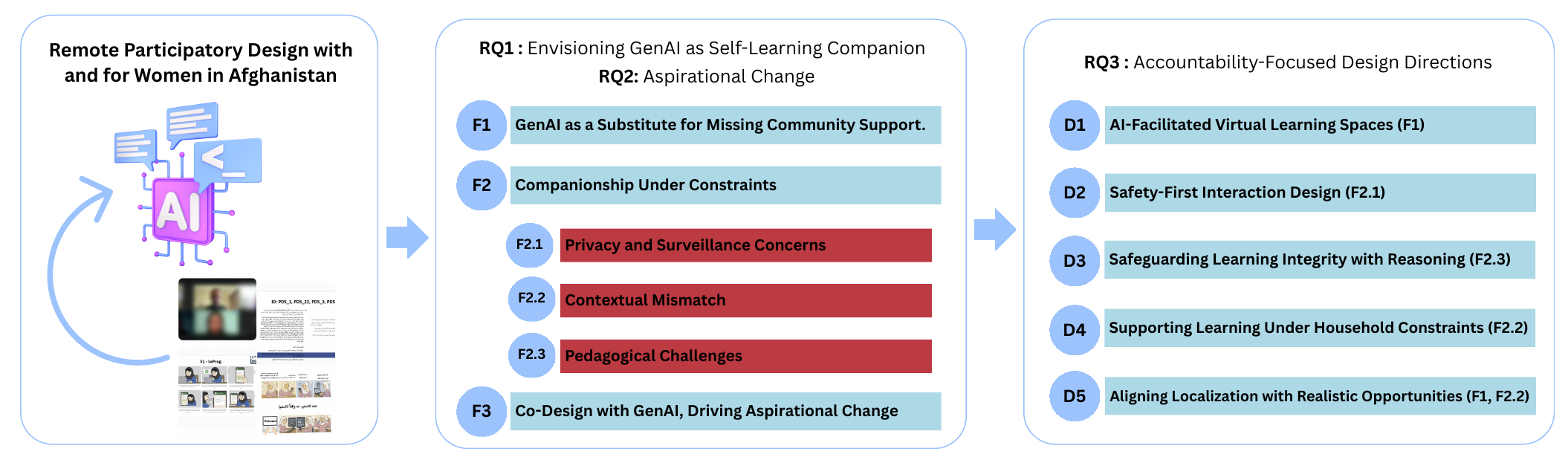}
  \caption{Overview of our participatory design study with women in Afghanistan to envision GenAI as a learning companion for online self-learning. The figure summarizes our findings (F1–F3) and accountability-focused design directions (D1-D5), building on participants' lived experiences where they are banned from formal education.}
  \Description{Overview of our participatory design study with women in Afghanistan to envision a GenAI learning companion for online self-directed learning. The figure summarizes our findings (F1–F3) and design directions (D1-D5), building on participants' lived experiences in a gender restrictive context where they are banned from formal education.}
  \label{fig:teaser}
\end{teaserfigure}

\maketitle

\section{Introduction}
Inclusive and quality education is a fundamental human right \cite{UN_GA_2015_2030Agenda, UNESCO_RightToEducation_2025}; however, women in Afghanistan have experienced a major contraction in access to formal education, particularly since 2021, pushing many to pursue learning through informal, online, and self-directed pathways instead \cite{behmanush2025online, sarwari2024alternative, unicef_unesco_education_2025}. These pathways increasingly rely on mobile phones and online platforms, but access remains uneven, as connectivity is often costly and unstable, devices may be shared, and learning resources in local languages remain limited \cite{behmanush2025online, osce_digital_transformation_2025, karimy2024review, unicef_unesco_education_2025}. In addition, women’s learning is shaped by household responsibilities, restricted mobility, and the loss of in-person peers, mentors, and institutional routines that would normally structure study \cite{adhikari2024learning, sultana2018design, osce_digital_transformation_2025, unicef_unesco_education_2025}. In this setting, digital tools are not simply optional supplements to formal education; they are among the few remaining infrastructures through which women can continue learning and pursue career aspirations.

Within this shift toward online self-learning, GenAI tools are increasingly framed as always-on educational support that can tutor, scaffold educational tasks, and guide learning trajectories \cite{chen2024gptutor, roe2024generative, jauhiainen2024generative}. In principle, GenAI can support self-learners by providing personalized feedback \cite{turner2025harnessing, roe2024generative}, generating tasks for practice, and helping learners organize their study trajectories \cite{sarsa2022automatic, wermelinger2023using, shihab2025effects, giray2025self}. However, prior work highlights risks central to accountability in learning with unreliable outputs, shallow engagement, over-reliance, and opaque data practices \cite{kasneci2023chatgpt, yan2024practical, holmes2023guidance, wang2024large}. For learners with limited prior educational opportunities, these risks are amplified by hallucinated content and unsupportive learning communities \cite{holmes2023guidance, shirzad2025ai, ibtasam2019my}. Given these challenges, we still know little about how women facing systematic educational exclusion envision GenAI as a learning companion that can responsibly support learning and employability under the constraints of privacy and surveillance. Moreover, although prior work has largely examined GenAI design and evaluation in formal educational settings \cite{zheng2024charting, pu2025can, ravi2025co}, comparatively little research has focused on designing with women in gender-restrictive contexts, where contextual mismatch and safety risks can render always-on support consequential. In such contexts, accountability is fundamentally about exposure: what traces GenAI use leaves and what risks those traces can trigger, as surfaced in our participatory design sessions.

Using participatory design (PD) as a methodological framework for designing inclusive EdTech and GenAI tools, this study addresses the following research questions:
\begin{itemize}
    \item \textbf{RQ1:} How do women in gender-restrictive contexts envision GenAI as a learning and employability companion, and what constraints shape its use?

    \item \textbf{RQ2:} How does participation in the participatory design of a GenAI learning companion associate with learners’ aspirations, including perceived agency and perceived avenues?

    \item \textbf{RQ3:} What accountability-focused design directions emerge for GenAI learning companions in gender-restrictive contexts?
\end{itemize}

This paper advances the understanding of accountable GenAI for education, particularly in gender-restrictive contexts, where privacy, surveillance, and infrastructural constraints fundamentally shape what constitutes safe and useful support. We highlight how women navigating systemic exclusion effectively use GenAI for self-learning while facing risks of exposure and misalignment, and underscore accountability gaps that are often overlooked in well-resourced contexts. Methodologically, we complement qualitative participatory design insights with pre-/and post-quantitative evidence that participatory, future-oriented design can shift learners’ reported aspirations, including perceived agency and avenues. Drawing on participants’ ideas, we articulate accountability-focused design directions for GenAI learning companions that (1) rebuild missing learning community support through AI-facilitated and level-based virtual learning spaces;  (2) adopt safety-first interaction designs by minimizing trace and maximizing learner control (e.g., anonymous use, rapid deletion); (3) support learning under household constraints with flexible and bilingual microlearning; (4) protect learning integrity by providing step-by-step reasoning over direct-answer interactions, and (5) balance localization to support effective learning with employability skill-building toward realistic working opportunities.

\section{Related Work}

\subsection{Safety, Accountability, and Learning Under Constraint}

As GenAI becomes more embedded in educational practice, questions of safety and accountability have become central to its responsible use \cite{cooper2022accountability, casper2024blackbox, dotan2024responsible, raji2020closing}. Prior work frames accountability as more than model performance: learners and institutions should be able to understand when GenAI is involved, what kinds of assistance it is providing, and what forms of oversight, contestation, and remedy are available when harm occurs \cite{cooper2022accountability, dotan2024responsible, metcalf2021aia, raji2020closing}. In education, these concerns are especially salient because GenAI can generate plausible but incorrect explanations, reproduce bias, and encourage forms of reliance that support task completion without necessarily supporting durable understanding \cite{kasneci2023chatgpt, yan2024practical, holmes2023guidance, wang2024large, bassner2025lessstress, barcaui2025cognitivecrutch}.

At the same time, work on situated and responsible AI argues that accountability cannot be understood only through transparency, correctness, or post-hoc governance \cite{katell2020situated, cooper2022accountability, casper2024blackbox}. In unequal and high-risk settings, accountability is also shaped by the social conditions under which systems are used. For learners studying under constrained conditions, this includes whether GenAI use leaves traces, what those traces may reveal, whether support is usable on shared or low-performance devices, and whether learners can control disclosure, retention, and visibility of their interactions \cite{holmes2023guidance, roe2024generative, karimy2024review, diberardino2023antiintentional}. In this sense, accountability is not only about explaining model behavior; it is also about reducing exposure and giving learners practical control over when, how, and how safely support can be used.

These concerns are particularly important in constrained self-learning environments, where educational access is shaped by unstable connectivity, limited local-language resources, device sharing, household responsibilities, and reduced access to peers, mentors, and institutional routines \cite{behmanush2025online, osce_digital_transformation_2025, karimy2024review, unicef_unesco_education_2025, adhikari2024learning, ibtasam2019my}. Under such conditions, learning support must be evaluated not only by answer quality, but also by whether it is contextually safe, realistic, and pedagogically appropriate. Educational studies further show that GenAI assistance can improve short-term performance while weakening independent problem-solving or confidence calibration when help is poorly calibrated \cite{bassner2025lessstress, barcaui2025cognitivecrutch, tan2024shaping}. Taken together, this literature suggests that accountability in constrained learning contexts must encompass learner control, privacy and exposure management, contextual safety, and support for genuine learning rather than quick answers alone.

\subsection{GenAI and Participatory Design in Education}
% Participatory Design in Restrictive Contexts
GenAI, particularly large language models, are now used in educational settings to provide conversational tutoring, generate examples and practice items, scaffold problem-solving, support writing and planning, and assist with programming tasks \cite{chen2024gptutor, zhang2024artificial, roe2024generative, guettala2024generative, jauhiainen2024generative}. A recurring argument is that such tools can approximate elements of individualized tutoring by offering responsive, on-demand support at scale \cite{bloom19842, turner2025harnessing, al2025exploring}. In parallel, emerging work explores how tailored GenAI interventions may support learners who have historically had limited access to high-quality instruction, particularly when integrated into broader ecosystems of human support and locally appropriate resources \cite{rahim2025harnessing, behmanush2025online, swaminathan2025generative}.

Participatory Design (PD) provides a complementary methodological foundation for shaping these tools with, rather than merely for, the people most affected by them \cite{muller1993participatory, disalvo2016participatory}. PD emphasizes mutual learning between researcher and participants, treats participants' values as design material, and explicitly attends to power and inequality, often with an orientation toward social justice and long-term structural change \cite{bjorgvinsson2010participatory, tuhkala2021systematic}. Within educational AI and GenAI, PD has been used to co-design learner-facing agents and creativity or learning-support tools, commonly through workshops and iterative prototyping paired with qualitative assessments of usability, perceived agency, and fit with learning goals \cite{buddemeyer2022unwritten, newman2024want, wang2025becoming, al2025exploring}.

Despite this growth, PD-for-GenAI in education still disproportionately centers formal institutions and comparatively well-resourced learner populations. Far fewer studies foreground women whose educational and employment opportunities are constrained by restrictive gender norms and socio-political instability, even though these contexts fundamentally shape both design constraints and what benefit means in practice. Moreover, while prior PD studies provide rich qualitative accounts of engagement and agency, the quantitative assessment of downstream outcomes, including shifts in aspirations and future envisioning with GenAI, remains rare \cite{wang2025becoming, harrington2020forgotten, laursen2025post}. This gap motivates approaches that combine participatory methods with measurable outcomes, while incorporating the accountability requirements of exposure control and safety constraints discussed above.

\section{Methodology}
This study is part of a broader research project \cite{behmanush2025supporting} that seeks to adapt GenAI to support marginalized learners in gender-restrictive and sociopolitically unstable contexts. All research activities were approved by our university’s Ethical Review Board (ERB).

\subsection{Participants and Recruitment}
We partnered with Code to Inspire (CTI)\footnote{\url{https://www.codetoinspire.org/}}, a coding school that supports girls in Afghanistan with online and programming education, to recruit participants for our PD study. To complement recruitment through the coding school and minimize the risk of overlooking less visible subgroups, we additionally used snowball sampling \cite{cohen2002research} in our recruitment process. In February 2025, we administered a brief bilingual (English–Persian) demographic survey that requested informed consent, recorded demographic characteristics, and documented participants’ access to and use of online learning and GenAI. The survey was shared through the coding school’s channels and via snowball referrals. In total, we received 143 responses (CTI: $n = 90$; snowball sampling: $n = 53$). Three participants did not provide consent to participate in the study.

After the demographic survey, we offered four separate 60-minute introductory sessions in April–May 2025 to familiarize the survey respondents with the PD process, answer questions, and encourage participation \cite{Nanyonjo2025GeJuSTA}. The sessions were conducted via Zoom in Persian by two researchers from Afghanistan (one female and one male) and provided a combined overview of PD, followed by an open Q\&A. No data were collected during these sessions. From the pool of consenting survey respondents who attended the introductory session, we randomly selected 20 participants to invite to the PD sessions. To help mitigate participation barriers \cite{Nanyonjo2025GeJuSTA}, we provided mobile data support to participants with connectivity challenges and engaged with a local gatekeeper to facilitate one participant’s involvement. The gatekeeper’s involvement was limited to the initial consent stage and served only to facilitate the participant’s attendance at the PD session. He did not attend the PD session, did not participate in the discussion, and did not set any rules or constraints on the participant’s contributions. We then conducted five PD sessions, each with four participants, in May–June 2025. We provided each PD participant with €20 in compensation, which we believe was appropriate given the session length and local economic conditions.

\subsection{PD Processes}
After the introductory sessions, we conducted five PD sessions on Zoom. Each session included four participants, was co-facilitated by two researchers, and lasted approximately three hours. We intentionally adopted a small-group format with two facilitators per session to reduce pressure, support safer participation, and encourage equitable turn-taking. Facilitators used careful prompting to encourage participation without pressuring participants to respond and periodically created space for quieter participants. The PD process was structured so that participants first articulated their own current practices, challenges, and desired forms of support in open discussion, and then used those reflections to critique and reshape researcher-prepared storyboard concepts. We used the storyboards as prompts for participants to identify misalignments, propose additions or removals, and articulate alternative interaction features and deployment choices grounded in their lived realities.

\begin{figure}[htbp]
  \centering
  \includegraphics[width=0.99\columnwidth]{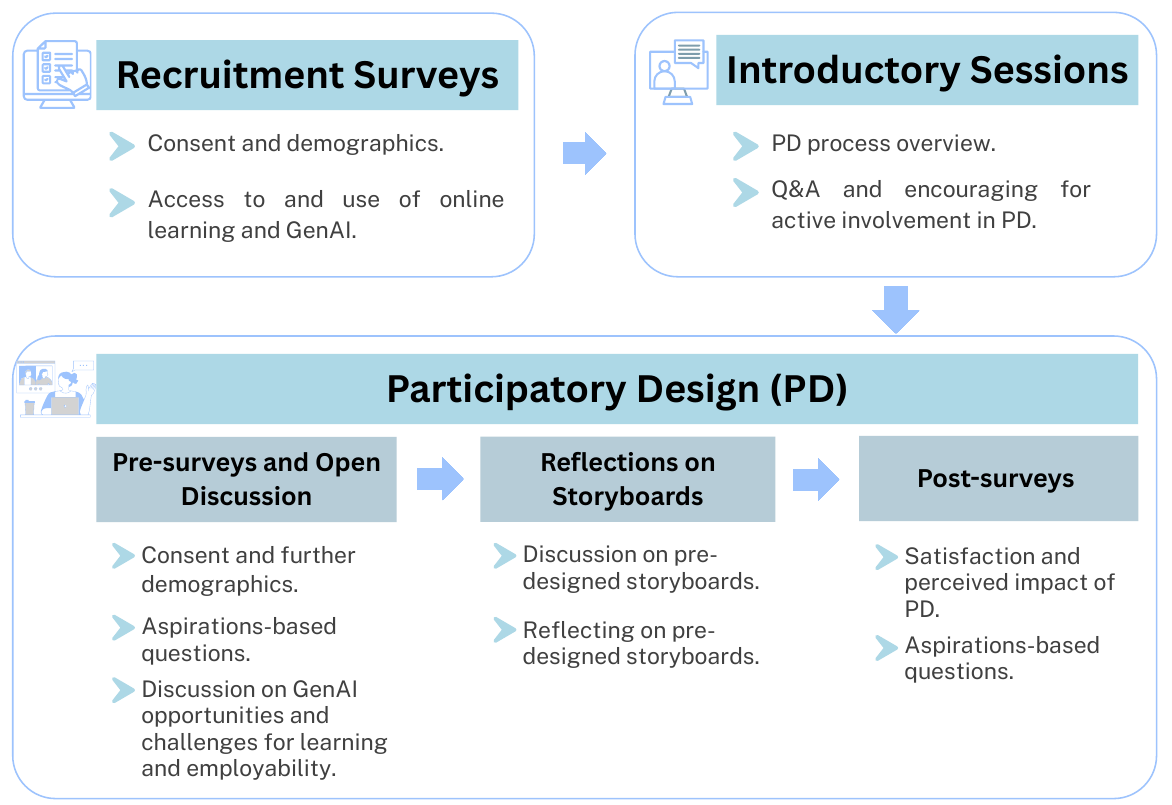}
  \Description{Overview of our study design. The figure illustrates the sequence of activities, beginning with recruitment surveys, followed by introductory sessions to familiarize participants with the participatory design (PD) process. This is then followed by PD sessions, including pre-PD surveys and open discussions, reflections on pre-designed storyboards, and concluding with post-surveys.}
  \caption{Overview of our study design. The figure illustrates the sequence of activities, beginning with recruitment surveys, followed by introductory sessions to familiarize participants with the participatory design (PD) process, then PD sessions including pre-PD surveys and open discussions, reflections on pre-designed storyboards, and concluding with post-surveys.}
  \label{fig:StudyDesign}
\end{figure}

% \begin{figure}[htbp]
%   \centering
%   \includegraphics[width=\columnwidth]{Figures/StudyDesign.pdf}
%   \Description{Overview of our study design. The figure illustrates the sequence of activities, beginning with recruitment surveys, followed by introductory sessions to familiarize participants with the participatory design (PD) process. This is then followed by PD sessions, including pre-PD surveys and open discussions, reflections on pre-designed storyboards, and concluding with post-surveys.}
%   \caption{Overview of our study design. The figure illustrates the sequence of activities, beginning with recruitment surveys, followed by introductory sessions to familiarize participants with the participatory design (PD) process, then PD sessions including pre-PD surveys and open discussions, reflections on pre-designed storyboards, and concluding with post-surveys.}
%   \label{fig:StudyDesign}
% \end{figure}

\subsubsection{Pre-Surveys and Open Discussion}
\label{P&OP}
PD sessions began with a pre-survey that recorded informed consent, further demographic information, and responses to an adapted aspiration-based scale \cite{behmanush2025hope, snyder1991will} to assess changes in participants’ aspirations associated with attending the PD sessions and envisioning their future with GenAI. The pre-survey was followed by an open discussion structured around four prompting questions: participants’ most recent questions posed to a GenAI tool, opportunities that generative AI offers for learning programming and supporting employability in their context, challenges with current tools, and the required features and capabilities of a GenAI companion to support learning programming and employability. The open discussion led to participant-generated scenarios that both surfaced challenges with existing tools and envisioned a companion with the desired capabilities. These participant-generated scenarios served as the baseline against which researchers' pre-designed storyboard concepts were later discussed, helping participants identify where the proposed companion aligned with, failed to reflect, or should be revised to better match their constraints, learning practices, and safety needs. At the end of each PD session, participants individually or in pairs translated their scenarios into storyboards. To reduce time pressure and support safer participation, they were allowed up to two hours after the session to finalize and submit their storyboard. An example storyboard received from participants is shown in Figure \ref{fig:Storyboardp} in Appendix D. Open discussions were audio-recorded with participants' consent, and facilitators took detailed notes.

\subsubsection{Storyboards}
\label{Strb}
Following the open discussion, participants evaluated four storyboards, each depicting a prospective GenAI companion for programming learning and employability support. Participants assessed the storyboards using four prompting questions: (i) their understanding of each storyboard and whether the scenarios reflected their own situations; (ii) which existing features were confusing or not helpful; (iii) what additions, removals, or changes they would recommend; and (iv) how the companion could be better adapted to their learning style and cultural context. Rather than treating this activity as a summative evaluation of predefined concepts, we used it as a generative design exercise for refining the companion.

\begin{figure*}[htbp]
  \centering
  \includegraphics[width=\textwidth]{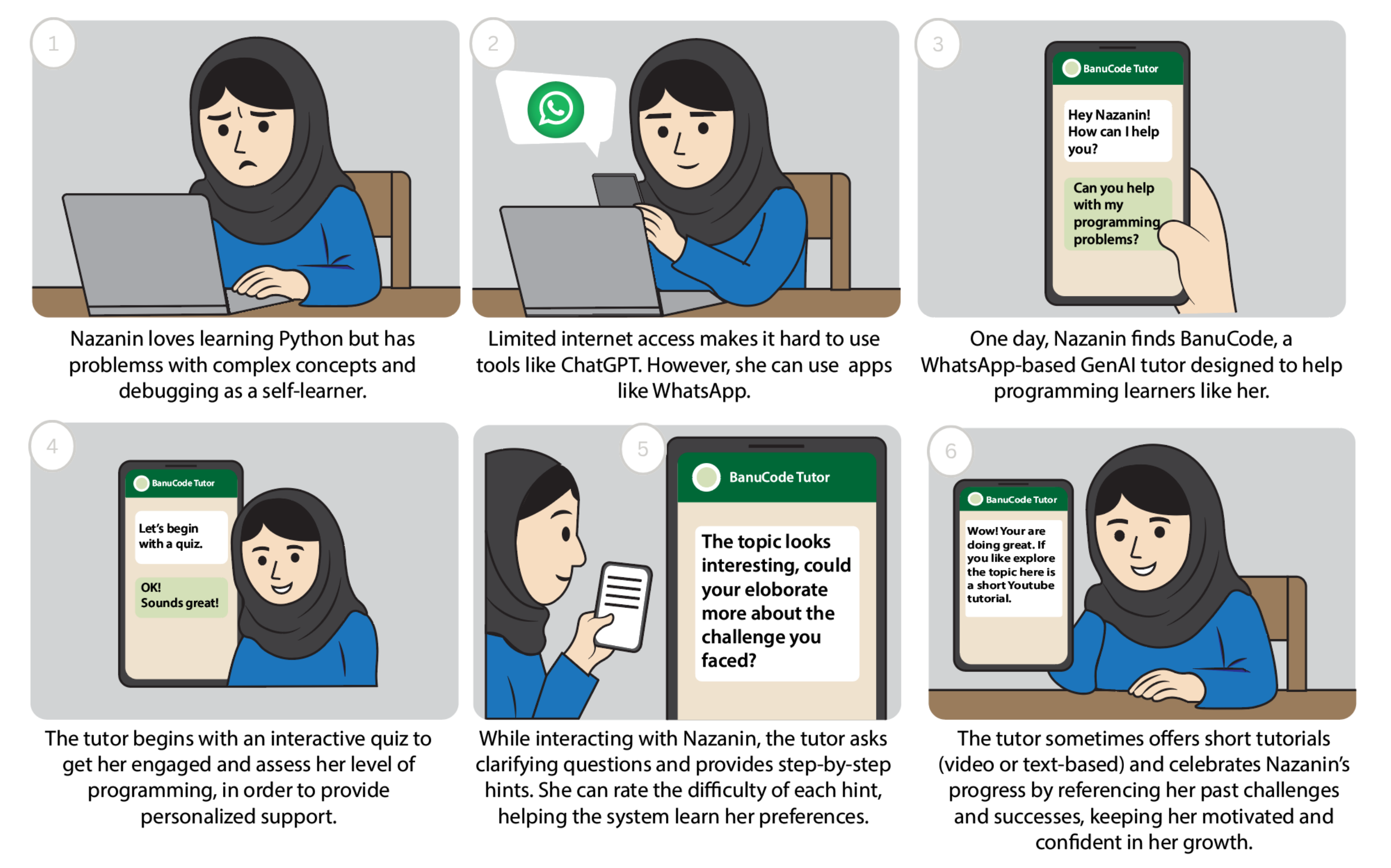}
  \Description{A sample storyboard used in PD.}
  \caption{An example pre-designed storyboard used in our participatory design sessions.}
  \label{fig:Storyboard}
\end{figure*}

During each session, facilitators documented participants’ responses to specific storyboard elements, including which features they wanted to retain, which they considered unrealistic or unsafe, and which they proposed as alternatives. We then examined these discussions across sessions to identify recurring design requirements while maintaining traceability between participants’ feedback and the resulting design directions. To strengthen traceability between the participatory design process and the final design directions, we documented storyboard feedback at the level of specific elements (e.g., interaction visibility) and recorded whether participants recommended retaining, removing, or modifying each element. For example, when participants commented on storyboard concepts involving peer- or platform-based support, they emphasized that interaction should not require identity-linked registration and should remain low-visibility on shared devices. These comments directly informed design implications related to anonymous participation, rapid trace removal, and user control over what remains visible or stored.

The four storyboards varied along two dimensions. First, they differed in the scope of support provided, encompassing not only programming assistance but also employability and soft-skill support for self-learners. Second, they differed in deployment modality, depicting the companion either as embedded within an existing application, implemented as a browser plugin, or provided as a standalone tool. The storyboard concepts were informed by prior work on conversational agents and AI-based support in low-resource learning environments (e.g., \cite{behmanush2025online, bringman2020students, angrist2022experimental, McNulty2025DigitalTeacherSupport, afoakwah2021dialling, wait2013reducing}). A sample storyboard used in the participatory design sessions is shown in Figure \ref{fig:Storyboard}.

\subsubsection{Post-Surveys}
A short post-survey was conducted to assess participants’ satisfaction and the perceived impact of the PD sessions. It included open-ended questions about whether the PDs were helpful and whether participants felt their ideas were heard, and what aspects they did not find helpful. Additionally, the adapted aspiration-based scale \cite{behmanush2025hope, snyder1991will} was reused in the post-survey to measure changes in participants’ aspirations after participating in the PDs and envisioning their future with GenAI.
\subsection{Data Collection}
We used a demographic survey administered prior to the PD sessions to collect background information and technology-use patterns. The survey included items on education level, age, preferred language of communication, employment status, and the frequency and use cases of online learning resources and generative AI. Only responses from participants who provided informed consent were included in the subsequent analyses. Demographic survey respondents’ ages ranged from 18 to 34 years (M = 23.4). Detailed demographics are presented in Table~\ref{tab:demographics} in Appendix B.

\begin{table}[h!]
  \caption{We conducted remote PD with 20 women from Afghanistan in 5 online sessions. The table below shows their demographics.}
  \label{tab:demographics1}
  \centering
  \begin{tabularx}{0.75\linewidth}{@{}lXr@{}}
    \toprule
    \textbf{Indicator} & \textbf{Response} & \textbf{N} \\
    \midrule
    \textit{Age}                    & 18--21                 & 6  \\
                          & 22--25                 & 6  \\
                          & 26--34                  & 8  \\
    \midrule
    \textit{Educational Background} & Computer Science (CS)  & 14 \\
                          & Non-CS                 & 6  \\
    \midrule
    \textit{Marital Status}         & Single                 & 15 \\
                          & Married                & 5  \\
    \midrule
    \textit{Learning Method}        & SDL                    & 17 \\
                          & SDL and Online Classes & 3  \\
    \bottomrule
  \end{tabularx}
\end{table}

During the PD sessions, we collected pre- and post-session survey responses, participant-generated scenarios and storyboards, and audio recordings of the open discussion and storyboard reflection sections. Across all sessions, participants, working individually or in pairs on scenario generation, produced 12 scenarios and 12 post-PD storyboard submissions, along with approximately 8 hours of audio recordings from discussions and reflections. The audio recordings were manually transcribed in the original language and then translated into English using machine translation. The first author subsequently reviewed the translated transcripts against the original-language transcripts, corrected translation errors, and refined wording where culturally specific or potentially ambiguous expressions risked losing their intended meaning. Participants in the PD sessions ranged in age from 19 to 34 years. Most identified as self-learners (85\%, $n = 17$), while the rest reported a combination of self-learning and structured online classes (15\%, $n = 3$). Demographic details for PD participants are provided in Table~\ref{tab:demographics1}.

\subsection{Data Analysis}
We used Python to clean the demographic, pre-PD, and post-PD survey data and to compute descriptive statistics.  We analyzed pre- and post-PD surveys to assess changes in participants’ aspirations. Statistical significance was tested with paired $t$-tests \cite{szczytko2018impacts,yang2022comparative}. In addition to $p$-values, we report a percentile-based effect size \cite{kraft2020interpreting}. For each scale, we first computed the median of the post-PD scores and then located this value within the pre-PD distribution, yielding the post-PD median percentile rank relative to the pre-PD distribution.

We qualitatively analyzed (1) participants' scenarios and post-PD storyboards and (2) transcripts of audio-recorded open discussions and storyboard-reflection activities. We treated the storyboards as visual narrative data and first examined them alongside the transcripts to identify recurring situations, constraints, desired features, and learning-support practices across the dataset \cite{truong2006storyboarding}. We then conducted an inductive, iterative thematic analysis \cite{braun2006using,braun2019reflecting} across the full dataset. One researcher led the initial familiarization and open coding of the transcripts and storyboard materials and developed a draft codebook grounded in emergent patterns in the data. This draft codebook was then iteratively reviewed with the broader research team to refine code definitions, merge overlapping codes, distinguish conceptually separate codes, and maintain traceability between raw participant accounts, intermediate codes, and higher-level themes. For example, participant requests for rapid deletion, use without registration, and non-identifying access were initially coded separately and later consolidated into a higher-level code family related to privacy, security, and anonymity. This code family contributed to the broader theme of safety, privacy, and trace control and informed the corresponding design direction of minimizing trace and maximizing learner control. After the team reached an agreement on the codebook, two coders independently applied it to the full dataset. Disagreements were resolved through discussion and, where needed, further refinement of code definitions before final theme generation. Inter-coder agreement yielded a Cohen's kappa of $\kappa = 0.72$, indicating substantial agreement \cite{landis1977measurement}. We then used the final codes and themes to develop our findings and articulate accountability-focused design directions for self-learners in gender-restrictive contexts. Appendix~A provides a concise codes-to-themes summary that maps the final themes to their associated code families.

\section{Findings}
In this section, we present our findings in four parts: (1) how marginalized women envision and value GenAI for online self-learning and employability; (2) the constraints that shape and often limit GenAI’s role in their context; (3) how participatory design and future-oriented envisioning with GenAI relate to shifts in participants’ reported aspirations; and (4) we synthesize an accountability-focused design direction for GenAI learning companions that better align with participants’ infrastructural, sociocultural, gendered, and linguistic realities.
\subsection{GenAI as a Partial Substitute for Missing Community Support} % Or GenAI as a Substitute for Missing Learning Communities
\label{F1}
Since the formal restrictions on women’s education and employment in Afghanistan in 2021, women and girls have increasingly shifted to online self-learning as an alternative pathway \cite{sarwari2024alternative, behmanush2025online}. Insights from our recruitment survey indicate that participants primarily utilize online learning to develop programming and other employability-related skills (See Figure \ref{fig:DemoData} in Appendix C). However, sustaining this learning is difficult due to limited and costly connectivity, weak community support, household responsibilities, and a shortage of learning resources in local languages \cite{behmanush2025online, osce_digital_transformation_2025, unicef_unesco_education_2025, sarwari2024alternative}. Even when access is available, uncertainty about how to initiate and structure self-study, without the guidance of instructors, peers, or institutional routines, further hinders the effective use of online learning.

Within this constrained learning environment, GenAI is already embedded in participants’ everyday study practices, not as an occasional supplement but as a frequently accessed support channel.  Recruitment survey data indicate intensive use of GenAI, as 56\% of respondents reported using GenAI daily, and an additional 33\% reported using it multiple times per week, indicating reliance patterns consistent with “always-available” assistance. Accordingly, participants in our PD did not characterize GenAI merely as an information source; rather, they repeatedly described it as a layered form of support that can partially compensate for gaps in peers, mentorship, and career guidance, shaping how they conceptualize GenAI as a learning partner within their self-learning environment.

\textbf{Peer-Like Presence.} The PD participants repeatedly emphasized GenAI’s constant availability as peer-like companionship during study sessions, describing it as “at the same level, sitting beside” (PD4\_OP; PD5\_3) them while they learn. This peer-like presence is envisioned as unbound by classroom schedules or physical infrastructure, but rather as an app installed on a mobile phone or laptop that remains “usable offline and [able to] update itself” (PD3\_1; PD3\_2; PD5\_OD) whenever connectivity is available, thereby providing support across time and place.
\begin{quote}
    \textit{“Another important feature is its [GenAI's] unlimited accessibility at any time and place ... or as a peer, who we can imagine sitting beside us and speaking with us at the same level.”} - PD4\_OP
\end{quote}

\textbf{Mentor-Like Support.} Beyond the GenAI peer role, a mentorship role also emerged across the PD sessions. Participants envisioned GenAI as a “knowledgeable and patient guide” (PD5\_OD; PD1\_3) that can walk them step by step through complex tasks, such as debugging or algorithm design, offering detailed explanations of complex topics in accessible formats, including text or voice, as needed. They also articulated a need for this mentor to provide “sustained, structured feedback over time” (PD2\_1; PD4\_2), including periodic summaries of what they have studied, where they performed well, and where they continue to struggle, so that they can “identify strengths and weaknesses” (PD2\_1) and adjust their learning efforts within the self-learning process.

\begin{center}
\begin{minipage}{0.75\linewidth}
\begin{ideabox}{GenAI compensates for the missing in-person learning community support.}
  \begin{itemize}[label=\ding{52}, leftmargin=1.2em]
    \item \textbf{Peer-like presence:} always-available study companion.
    \item \textbf{Mentor-like support:} step-by-step explanations and feedback.
    \item \textbf{Employability guidance:} practice soft skills and career preparation support.
  \end{itemize}
\end{ideabox}
\end{minipage}
\end{center}

\textbf{Employability Guidance.} Participants linked employability challenges to the loss of in-person classes and the realities of self-learning, emphasizing that without classrooms and peers, they have fewer routine opportunities to develop soft skills. In this context, GenAI was envisioned as compensating for what self-learning cannot easily provide; for instance, helping them practice “time management or communication and professional interaction such as email writing” (PD4\_2; PD4\_4; PD5\_OD) or using role-play to “simulate a discussion between” them as a freelancer and “clients in Upwork” (PD3\_2; PD4\_2). Importantly, they framed this employability support as concrete and pathway-specific rather than generic advice, including guidance oriented to platforms like Upwork and Fiverr, assistance with “proposal writing” (PD4\_4), and preparation for “technical and coding interviews” (PD4\_4; PD5\_OD), alongside information about relevant remote job markets (PD4\_4; PD5\_OD). 

Taken together, this peer-like presence, mentorship, and employability guidance position GenAI as an accompanying tool for these self-learners in the absence of in-person learning communities. However, this companionship comes with limitations that constrain its usability and effectiveness.

\subsection{Companionship Under Constraints: Privacy, Surveillance, and Contextual Mismatch}
\label{F2}
\subsubsection{Privacy and Surveillance Concerns}
While the PD participants envision GenAI as their learning companion, they also identify constraints and tensions in its use.  Participants highlighted privacy as a key concern for the use of GenAI as an online learning companion. They explicitly framed “the protection of privacy as a major challenge” (PD2\_OD) associated with these tools. They stressed that even minor privacy breaches could have severe consequences for them, considering the restrictive gender norms in the context. 
\begin{quote}
    \textit{“[...] for girls, protecting personal information and privacy is a critical issue. Even the smallest problem in this regard could cause lots of difficulties for us.”} - PD2\_3
\end{quote}
Privacy was understood not only in technical terms, but also as deeply entangled with gendered and social risks. Participants emphasized that “personal security and protection of private information is extremely important” (PD3\_3) and therefore any educational technology must be taken seriously on this front. 

Closely related to privacy, participants highlighted surveillance by household monitoring or even authorities as another constraint that shapes what they can ask, when they can ask it, and how visible their learning must remain. They described living under “excessive restrictions” (PD2\_2; PD3\_OD) where “even asking for information from this tool might cause problems” (PD2\_2). Participants did not discuss surveillance in the abstract; instead, they described concrete, everyday risks that they faced. Given their reliance on “shared devices” (PD3\_2), even in some cases, revealing learning content to other family members can cause serious problems. Participants emphasize that in a context where “use of mobile phones sometimes causes problems” (PD5\_4), the tool’s suggestion of a “dating application as a project” (PD2\_1) does not align with local norms and could lead to tension within families. These highlight that for learners in such patriarchal contexts, the practical realities of device access, monitoring by gatekeepers, and not aligning with local norms can carry significant social risks.
\subsubsection{Contextual Mismatch}
\label{F2.2}
In addition to privacy and surveillance challenges, participants consistently reported a persistent mismatch between existing GenAI outputs and their local educational, cultural, and labor market contexts. Focusing on employability, participants noted that generated content “often does not match local needs or the skills required” (PD4\_OD), instead focusing on professions relevant to developed contexts. Contextual misalignment was not only noted in employability but also in everyday learning advice and suggestions. Participants described recommendations for GenAI tools, such as studying in a mixed-gender public space or working from an internet café, which they believe are unrealistic in their context. 
\begin{quote}
    \textit{“[...] or recommends going outside the home to carry out an activity in a place like an internet café. These kinds of suggestions are not suitable for us as girls in Afghanistan.”} - PD1\_1
\end{quote}

The contextual mismatch reported by participants extended into the technical sphere, where GenAI outputs were often incompatible with their constrained resource availability. Learners expressed concern that GenAI-generated code sometimes requires “high-capacity processors and powerful devices” (PD4\_1, PD4\_OD, PD5\_4). This was compounded by their direct experience that the tools sometimes generated codes that “cannot be executed” (PD4\_OD), which reinforced their perception that the tools do not consider their context and circumstances while supporting them.

\begin{center}
\begin{minipage}{0.90\linewidth}
\begin{ideabox} [colback=red!5,colbacktitle=red!60!black]{GenAI companionship is constrained by privacy, surveillance, and contextual mismatch.}
  \begin{itemize}[label=\ding{55}, leftmargin=1.2em]
    \item \textbf{Privacy \& surveillance:} Small breaches and monitoring raise real risks.
    \item \textbf{Contextual mismatch:} Suggestions, code, and language support often don’t fit contextual realities.
    \item \textbf{Pedagogical risk:} Direct answers create an illusion of learning progress.
  \end{itemize}
\end{ideabox}
\end{minipage}
\end{center}

Furthermore, participants described how the tools are linguistically misaligned with their everyday realities, which “challenges their learning process” (PD2\_4; PD4\_1; PD1\_2), such as the rising tension between the need for native language support for immediate learning and the need for English proficiency for employability. While using native language makes concepts easier to grasp, learners cautioned that if they "focus too much on native" (PD2\_1) while lacking skills in English, they end up limiting their "own opportunities" (PD2\_1; PD1\_2) for jobs in the broader freelancing market. They expressed that current existing tools fail to adequately support learners in this context and rarely provide structured pathways that enable them to benefit from native-language explanations while systematically developing the English skills required for real-world work opportunities.

\subsubsection{Pedagogical Challenges}
\label{F3.2}
Pedagogical concerns, particularly the risk that GenAI might create an “illusion of learning” (PD2\_OD; PD4\_1; PD1\_3), highlight that “relying too much on artificial intelligence” (PD2\_OD; PD4\_1) may discourage them from learning new subjects and prevent the real growth of their skills. Several participants described how receiving complete solutions from existing tools undermined their thinking processes. 
\begin{quote}
    \textit{“[...] from GenAI tools in my learning process, because I believe it takes away my thinking ability and weakens my critical thinking.”} - PD2\_1
\end{quote}
PD participants reported that, through extensive use of these tools, they realized they were not “actually learning anything”(PD1\_3) but instead had “the illusion of learning”(PD1\_3). Importantly, they do not reject GenAI on this basis; rather, they raise this concern so that “learners become aware”(PD2\_1) of this risk and technologists can develop strategies to address it. They emphasize that this should be taken seriously, as in formal learning settings, it may be mitigated by the fear of exams, whereas in their non-formal learning context, it can persist unnoticed, gradually undermining learning and skill development.

\subsection{Participatory Design to Envision the Future with GenAI: Driving Aspirational Change}
We examined how learning aspirations of marginalized women changed by participating in PD and envisioning the future with GenAI using an adapted aspiration-based scale \cite{behmanush2025hope, snyder1991will}. The scale captures learners’ aspirations as perceived long-term goals that extend beyond their current circumstances \cite{toyama2018needs, cannanure2022we, cannanure2020m}. It includes two subscales: Agency, which represents their perception of ability to achieve those goals \cite{toyama2018needs, lybbert2017hope, cannanure2022we, cannanure2020m}, and Avenue, which refers to the viable paths they see for reaching those goals \cite{toyama2018needs, lybbert2017hope, cannanure2022we, cannanure2020m}. Participants reported higher scores on the aspirations measure after the PD sessions, as shown in Table~\ref{tab:pd_results}. Aspiration scores increased from \(M = 31.5\) (\(SD = 3.3\)) at pre-PD to \(M = 34.8\) (\(SD = 3.7\)) at post-PD. The result from the paired \(t\)-test showed that this difference was statistically significant, \(p = .01\). This suggests that, following the PD and envisioning the future with GenAI, learners articulated more long-term goals for their education and perceived these goals more positively within their circumstances.   

\begin{table}[H]
  \small
  \centering
  \setlength{\tabcolsep}{3pt}
  \caption{\small Pre- and post-PD scores with $t$-tests.}
  \label{tab:pd_results}
  \begin{tabular}{lrrrrrr}
    \toprule
    & \multicolumn{2}{c}{Pre-PD} & \multicolumn{2}{c}{Post-PD} & \multicolumn{2}{c}{ $t$-test} \\
    \cmidrule(lr){2-3}\cmidrule(lr){4-5}\cmidrule(lr){6-7}
    Scale & Mean & SD & Mean & SD & $t$ & $p$ \\
    \midrule
    Agency      & 15.7 & 2.0 & 17.3 & 2.1 & 2.2 & \textbf{0.03} \\
    Avenue      & 15.8 & 1.5 & 17.5 & 1.9 & 2.9 & \textbf{0.01} \\
    Aspiration  & 31.5 & 3.3 & 34.8 & 3.7 & 2.7 & \textbf{0.01} \\
    \bottomrule
  \end{tabular}
\end{table}

Looking at the sub-scales, agency scores increased from \(M = 15.7\) (\(SD = 2.0\)) at pre-PD to \(M = 17.3\) (\(SD = 2.1\)) at post-PD, \(p = .03\), indicating a statistically significant strengthening of participants' belief in their ability to move toward their goals. Avenue scores likewise improved from \(M = 15.8\) (\(SD = 1.5\)) to \(M = 17.5\) (\(SD = 1.9\)), \( p = .01\), providing evidence that learners left the PD with a sense of more paths for achieving their aspirations.

% \begin{figure}[H]
%   \centering
%   \includegraphics[width=\columnwidth]{Figures/Percentile-Change.png}
%   \Description{A description of the Percentile-Change figure.}
%   \caption{Participants’ median percentile ranks on Aspiration, Avenue, and Agency increased from the 50th percentile before the PD to about the 78th–80th percentile afterward.}
%   \label{fig:Percentile}
% \end{figure}
A complementary percentile-based effect size analysis provided additional insight into the scale of these changes. For the aspirations, the post-PD median corresponded to a percentile rank around 77th of the pre-PD distribution, a shift that falls within the range of a substantial improvement in practical terms. Similarly, for the Agency and Avenue sub-scales, the post-PD medians were located around the 80th percentile of their respective pre-PD distributions, which likewise qualify as substantial changes. These patterns align with a practically meaningful effect of the PD on participants' aspirations, Agency, and Avenue.

Additionally, through the post-survey, PD participants valued the session and highlighted these sessions as an opportunity to have their voices heard (\(n = 18\)), sharing their ideas through discussion (\(n = 11\)), opportunity for teamwork and collaboration (\(n = 7\)), and a supportive approach to their critical thinking and gaining new insights (\(n = 5\)).

\subsection{Designing Accountable GenAI Companion: Toward Trusted Support in Constrained Learning Environments} 
Designing accountable GenAI learning companions in gender-restrictive contexts is not only about providing assistance, but also about fostering trust among learners, where trust does not imply uncritical confidence, but rather a warranted reliance. This requires that learners can anticipate the companion’s behavior and reduce exposure to harm when privacy, surveillance, and contextual mismatch make even minor failures consequential. Building on the PD participants’ ideas, this section translates these requirements into accountability-focused design directions.

\subsubsection{AI-Facilitated Virtual Learning Spaces}
During PD sessions, participants envisioned the GenAI companion as a gateway to within-platform peer communities that could compensate for missing learning interactions. They contrasted their circumstances with less constrained settings where learners can “participate in [learning] community activities” (PD2\_4; PD3\_2), ask questions, and develop communication and collaboration skills, emphasizing that the absence of accessible in-person communities limits opportunities for soft-skill development. To address this gap, participants proposed that the companion host “a shared chat room among users of the same level” (PD5\_3), where learners share experiences and carry out group activities anonymously, with the companion providing guidance and moderation. Taken together, these accounts motivate GenAI designs that foreground AI-facilitated, level-based virtual learning spaces and structured peer collaboration, rather than relying solely on isolated one-to-one companions. Consistent with our finding that participants framed GenAI as a substitute learning companion (Section~\ref{F1}) and resonant with prior work on conversational tutoring and learner-centered agents (e.g., \cite{anderson2014-engaging, buddemeyer2022unwritten, newman2024want, behmanush2025online, cannanure2025enabling}), this positions GenAI as a facilitator that enables collaborative practice and the development of technical and soft skills for learners who lack access to in-person communities and classrooms.
\subsubsection{Safety-First Interaction Design}
Participants treated privacy and safety as the precondition for GenAI companionship, framing privacy not as a preference but as a high-stakes issue where “even a very small breach can cause a major” (PD3\_3) safety problem. Accordingly, they articulated interaction requirements that minimize trace and maximize learner control over what is stored, shown, and left behind on devices. Concretely, participants asked for “a simple and quick option to delete anything that has been used” (PD3\_3; PD2\_2), alongside use “without registration” (PD2\_2; PD2\_1; PD4\_4; PD5\_4) or through non-identifying IDs. These requests position privacy-by-design not as back-end compliance, but as an interaction-level guarantee that supports safe use in the presence of surveillance constraints (Section~\ref{F2}). Safety-first design also required boundary-aware and contextually safe support, because harm can arise not only from retained data but from suggestions that violate social constraints. Participants wanted the companion to respect “the boundaries that are non-negotiable” (PD4\_3) and to avoid guidance that could place them at risk. For example, recommendations to do tasks “in a shared environment alongside men” (PD1\_1) or project ideas (e.g., a “dating application”) that could trigger risks if seen by family members. 

Taken together, these accounts motivate safety-first interaction design as a combined commitment to trace minimization, learner agency, and boundary-conditioned generation which consistent with research showing how learning and technology use are continually negotiated under monitoring and gendered risk \cite{adhikari2024learning, sultana2018design, karusala2019privacy, varanasi2022feeling}, and aligned with calls for contextually safe and equitable learning technologies \cite{sultana2018design, shrestha2025bringing, raza2022fostering, melake2025whatworks}.

\subsubsection{Safeguarding Learning Integrity Through Step-by-Step Reasoning}
Participants did not object to GenAI because it can respond quickly; they worried that direct-answer support can undermine learning integrity when no teacher or peer is available to surface misconceptions or repair fragile understanding (Section \ref{F3.2}). They talk about a turning point after heavy use, where they realized they were not “actually learning anything” (PD1\_3) and were left with the illusion of learning progress, forgetting how to structure programs and even how to debug code. Some participants emphasized that providing complete answers “takes away” (PD1\_3; PD2\_1; PD1\_OD) their thinking ability and weakens their critical thinking. Consistent with prior concerns about shallow engagement and over-reliance in educational GenAI use \cite{kasneci2023chatgpt, yan2024practical, harvey2025don, wang2024large}, participants’ integrity agenda centered on interactions that the companion should privilege reasoning by prompting learners to attempt intermediate steps and by revealing support gradually, rather than collapsing tasks into final answers. They also emphasized learner control support, considering the time pressure and difficulty of enabling quick clarification when needed, while preserving deeper practice on complex problems, thereby operationalizing the companion as reasoning support rather than solution delivery.

\subsubsection{Supporting Learning Under Household Constraints}
Considering the learners' household responsibilities, limited study time, and language and infrastructural barriers (Section \ref{F2}), participants emphasized that GenAI support must be usable in fragments rather than merely available. They described having “dozens of family responsibilities” (PD2\_3; PD1\_2) and proposed an “environment for microlearning” (PD2\_3; PD5\_4; PD4\_4; PD2\_1) that fits 10–15 minute sessions, preserves state, and enables resumption with lightweight recaps when uninterrupted study is not feasible. Infrastructural limits further shaped accountability requirements (Section \ref{F2.2}), which participants asked for offline and “weak internet” (PD2\_OD; PD1\_OD) resilience, including offline access with occasional updates instead of continuous connectivity, and they cautioned that outputs (e.g., code) must run on low-spec devices to avoid becoming an additional barrier. They also framed bilingual support as a progression rather than a binary choice, requesting bilingual support for local-language explanations using familiar terms while keeping English visible to avoid “limit[ing] opportunities”  (PD2\_1; PD4\_4) tied to remote work. Together, these proposals suggest that accountable companions should couple resumable microlearning with low-bandwidth/offline operation, device-appropriate outputs, multimodal explanations beyond “text-only interaction” (PD4\_3), and bilingual support that bridges toward employability-relevant English, extending prior work on low-resource learning support and the shortcomings of generic, English-dominant materials in constrained contexts \cite{nigatu2024low, behmanush2025online, melake2025whatworks}.
\subsubsection{Aligning Localized Support With Realistic Work Opportunities}
Attention to locally grounded, project-based learning emerged as a core expectation for GenAI support. Participants critiqued existing tools for offering generic examples and career suggestions that overlook their “location, family conditions, and permission to travel” (PD1\_3), and instead emphasized small, contextually meaningful projects that build “self-confidence” (PD3\_1) and help the “knowledge stay” (PD2\_1; PD3\_4) in their minds. They framed this kind of localization as both pedagogically necessary and professionally consequential that learning activities should feel safe and usable in their everyday environment while also producing artifacts and skills that can translate to remote working opportunities. Accordingly, participants requested interactional scaffolds that “simulate a discussion” (PD3\_2; PD4\_4) between a freelancer and a client, alongside practical preparation for remote-work platforms such as Upwork and Fiverr, given the constraints on in-person employment. This emphasis responds directly to contextual irrelevance (Section~\ref{F2.2}) and operationalizes the companion as both a substitute for missing learning support and a pathway-oriented employability guide (Section~\ref{F1}), aligning with broader calls to design context-sensitive systems that respect constraints while expanding realistic opportunities for women (e.g., \cite{sultana2018design, shrestha2025bringing, raza2022fostering, melake2025whatworks}).

\section{Discussion}

Our findings suggest that, in restrictive learning environments, the central question for educational GenAI is not only whether it generates correct answers, but whether it affords \emph{warranted reliance}: support that learners can use safely, pedagogically, and under their actual conditions. This reframes educational GenAI as a socio-technical support system whose usefulness depends on social scaffolding, exposure control, contextual and pedagogical fit, and future-oriented participatory design.

\subsection{Beyond Direct Answers: GenAI as a Partial Stand-In for Missing Learning Communities}

Prior work on GenAI in education often evaluates support in terms of explanation quality, correctness, and adaptivity \cite{chen2024gptutor, roe2024generative, turner2025harnessing}. Our findings show that these criteria are not sufficient in contexts where learners have lost access to peers, mentors, and institutional routines. Participants valued GenAI not mainly for speed, but for continuity: someone to study with, guidance that accumulates over time, and help connecting learning to work. In this sense, the relevant unit of support is not a single answer, but an ongoing relationship to study.

This shifts the evaluation of educational GenAI in restrictive environments. GenAI systems should be judged not only by whether they answer well, but also by whether they help learners receive longitudinal feedback and recover some of the functions that formal education would ordinarily provide. At the same time, our findings point to an important tension: the more GenAI approximates community, the more carefully visibility, identity, and trace retention must be managed. Community-like support may be valuable, but only when anonymity, minimal retention, and user control are treated as first-order design requirements rather than optional safeguards.

\subsection{Accountability as Exposure Control}

In high-risk learning environments, accountability cannot be reduced to fairness, transparency, or auditability alone \cite{cooper2022accountability, raji2020closing, dotan2024responsible, metcalf2021aia}. Our findings show that it must also address \emph{exposure}: what remains on the device, what others can infer from use, and whether otherwise plausible suggestions become unsafe in local conditions. This broadens harm from incorrect outputs to include \emph{trace harms}, such as saved histories or registration footprints, and \emph{boundary harms}, where seemingly reasonable recommendations become risky because they exceed local social constraints \cite{katell2020situated}.

The implication is practical. Exposure control must be built into the interaction layer rather than treated only as a back-end privacy principle. Anonymous or temporary access, minimal history retention, rapid deletion, and user-steerable boundaries are not peripheral features in this setting; they are conditions of safe use. Accountable GenAI, then, is not only about preventing harmful outputs but also about giving learners meaningful control over when support becomes visible, durable, and risky.

\subsection{Localization as Situated Pedagogy and Opportunity Alignment}

Our findings suggest that guidance may be factually correct yet still unusable if it assumes uninterrupted study time, private devices, stable connectivity, or futures that are not realistically available. Localization in this setting, therefore, cannot be reduced to translation only. It is better understood as the production of support that learners can actually act on under their everyday constraints \cite{katell2020situated, behmanush2025online, holmes2023guidance}.

This requires alignment at three levels. First, \emph{pedagogical alignment}: support should fit fragmented study through resumable microlearning, lightweight recaps, and reasoning-first assistance rather than immediate solution delivery \cite{bassner2025lessstress, barcaui2025cognitivecrutch}. Second, \emph{infrastructural alignment}: generated code, tools, and workflows should assume low-spec devices and weak connectivity by default. Third,  \emph{opportunity alignment}: examples, projects, and career guidance should connect to realistic pathways such as remote freelancing, proposal writing, portfolio building, client communication, and interview preparation rather than generic advice detached from local constraints \cite{osce_digital_transformation_2025, rahim2025harnessing}. 

\subsection{Envisioning the Future with GenAI: PD Beyond Idea Generation}

Following the participatory design (PD) sessions, participants reported statistically significant increases in aspirations, perceived agency, and perceived avenues toward long-term goals. Although these pre/post shifts should not be read as evidence of durable, long-term change, they suggest that future-oriented participation around GenAI can have value beyond mere requirement elicitation. This interpretation is consistent with prior work showing that participatory processes can strengthen agency, self-efficacy, and reflective engagement \cite{harrington2020forgotten, laursen2025post, wang2025becoming}. It also resonates with foundational PD arguments that participation matters not only because it improves artifacts, but because it makes participants’ values, constraints, and aspirations consequential in shaping technology \cite{muller1993participatory, disalvo2016participatory, bjorgvinsson2010participatory}.

In this context, the significance of PD lies not only in improving the contextual fit of a proposed GenAI system but also in creating a rare space for collective reflection under conditions where educational and professional futures are otherwise constrained. Participants described the sessions as opportunities to be heard, exchange perspectives, and critically examine what GenAI could and could not do for them. This suggests that PD contributed not merely by eliciting preferences, but by helping participants articulate more concrete and plausible trajectories toward learning and employability. Seen this way, PD in accountable AI should be understood not only as a design methodology, but also as a process through which participants’ constraints, aspirations, and imagined futures become materially relevant to the direction of technological design.

\section{Conclusion}
This study examines how women banned from formal education in Afghanistan use GenAI for online self-learning and employability. Participants framed GenAI less as an information source and more as a peer-like, mentor-like, and career-guidance companion that partially substitutes for missing learning communities. Yet they emphasized that companionship can also increase exposure: traces on shared devices, surveillance, and contextually unsafe suggestions that can make learning a risky endeavor. Envisioning future GenAI-companions through participatory design was associated with significant increases in aspirations, perceived agency, and perceived avenues, suggesting that participatory, future-oriented methods can shift not only design requirements but also learners’ sense of perceived opportunities. We therefore frame accountable GenAI support as exposure control in this context by minimizing trace by default, maximizing learner control (e.g., anonymous use, rapid deletion), and constraining outputs to non-negotiable social boundaries. From this perspective, educational GenAI should be evaluated by warranted reliance: whether it helps learners sustain routines and pursue realistic opportunities without increasing harm, not only by the quality of its outputs.

\begin{acks}
IW and VC  are supported by funding from the Alexander von Humboldt Foundation and its founder, the Federal Ministry of Education and Research (Bundesministerium für Bildung und Forschung). We additionally thank Code to Inspire (CTI) for their valuable support with participant recruitment for this study.
\end{acks}

\section*{Positionality and Ethical Considerations}
Our research team is based in Germany and Afghanistan and brings experience in GenAI for education, human--computer interaction for development, and digital gender gaps. Some team members have themselves experienced restrictions on education and employability, which informed our sensitivity to power imbalances between researchers and participants.

This study was approved by our university’s Ethical Review Board (ERB). Given the gender-restrictive and sociopolitically unstable context, we implemented additional safeguards, including multi-stage informed consent, voluntary participation with the option to withdraw at any time, and mobile data support for participants who requested it. We minimized the collection of identifying information, deleted audio recordings after transcription, and restricted access to study materials to the research team in line with ERB-approved procedures.

\section*{Generative AI Usage}
We used ChatGPT 5 to suggest grammar and clarity edits on author-written text. All scientific content, analysis, results, and interpretations were produced by the authors; no GenAI tools were used to write the actual text of the findings or analysis.

% \clearpage

\bibliographystyle{ACM-Reference-Format}
\bibliography{ref}

\appendix
\switchonecolumn
% \appendix
% \onecolumn
% Keep numbering continuous and numeric
\renewcommand{\thetable}{\arabic{table}}
\renewcommand{\thefigure}{\arabic{figure}}

\section*{Appendix A: Final Themes and Associated Codes}
\label{app:themes-codes}

\begin{center}
\begin{minipage}{\textwidth}
\centering
\small
\refstepcounter{table}\label{tab:themes-codes}
\textbf{Table \thetable.} Mapping of the final themes to the associated codes derived from the thematic analysis.\par
\vspace{2pt}

\setlength{\tabcolsep}{4pt}
\renewcommand{\arraystretch}{0.99}

\begin{tabular}{@{}p{0.44\textwidth} p{0.56\textwidth}@{}}
\hline\hline
\textbf{Themes} & \textbf{Codes} \\
\hline

\textbf{GenAI as a learning and employability companion} &
\begin{tabular}[t]{@{}p{\linewidth}@{}}
AI as a Necessary Alternative \\
Challenge: Barriers to Independent Learning \\
Challenge: Career Path Uncertainty \& Lack of Experience \\
Challenge: Interview \& Work-Readiness Anxiety \\
Feature: Career Readiness \& Employability Guidance \\
Feature: Soft Skills \& Employability Support \\
Feature: Freelancing Skills Support \\
Feature: Tutor Feedback Mechanisms \\
Feature: Motivational Learning Support \\
Solution: Interview Preparation
\end{tabular}
\\[-2pt]
\hline

\textbf{Safety, privacy, and trace control} &
\begin{tabular}[t]{@{}p{\linewidth}@{}}
Challenge: Privacy, Security \& Data Ethics \\
Feature: Privacy, Security \& Anonymity \\
Feature: Safe Virtual Learning Community \\
Cultural Adaptation: Localized \& Sensitive Content \\
Feature: Offline Access to Chat Histories
\end{tabular}
\\[-2pt]
\hline

\textbf{Resource-constrained access and usability} &
\begin{tabular}[t]{@{}p{\linewidth}@{}}
Challenge: Access \& Infrastructure Limitations \\
Challenge: Platform Limitations (WhatsApp) \\
Feature: Offline \& Low-Bandwidth Functionality \\
Feature: Low-Resource Device Support \\
Feature: Need for Simple Prompting \\
Feature: Multimedia Support \\
Feature: Accessibility via Transcripts \\
Learning Style: Micro-Learning (Short Sessions) \\
Learning Style: Multimodal Learning (Text, Visual, Audio, Video)
\end{tabular}
\\[-2pt]
\hline

\textbf{Learning quality, scaffolding, and evaluation} &
\begin{tabular}[t]{@{}p{\linewidth}@{}}
Challenge: Illusion of Learning (Over-Reliance) \\
Challenge: Inaccurate or Low-Quality Responses \\
Challenge: Programming Difficulties and Educational Disruption \\
Feature: Follow-up Clarifying Questions \\
Feature: Integrated Code Execution \& Debugging \\
Feature: Learning Analytics \& Progress Tracking \\
Feature: Need for Clarity in Assessment Process \\
Feature: Need for Simplicity \& Detail \\
Feature: Adaptation to Learning Style \\
Learning Style: Contextualized Example-Based Learning \\
Solution: Step-by-Step Support \\
Solution: Personalized Learning \& Knowledge Checks \\
Use Case: Backend Programming \& Design Support
\end{tabular}
\\[-2pt]
\hline

\textbf{Language and cultural fit and contextual relevance} &
\begin{tabular}[t]{@{}p{\linewidth}@{}}
Challenge: Educational Restrictions \& Language Barriers \\
Challenge: Inaccurate Translation \\
Challenge: Lack of Context Awareness \\
Feature: Bilingual Support \\
Feature: Age-Appropriate Content \\
Solution: Multilingual \& Terminology Support \\
Solution: Native-Language \& Personalized Learning
\end{tabular}
\\[-2pt]
\hline

\textbf{Practical productivity and support use cases} &
\begin{tabular}[t]{@{}p{\linewidth}@{}}
Challenge: Data Collection Methods \\
Solution: Research \& Data Collection Support \\
Use Case: Everyday Tasks (Health, Time Conversion) \\
Use Case: Idea Generation \\
Use Case: Summarization \& Project Q\&A
\end{tabular}
\\[-2pt]
\hline\hline
\end{tabular}
\end{minipage}
\end{center}

\section*{Appendix B: Demographics of Recruitment Survey Respondents}
\label{app:demographics}

\begin{center}
\begin{minipage}{0.65\textwidth}
\small
\refstepcounter{table}\label{tab:demographics}
\noindent\parbox{\textwidth}{\raggedright
\textbf{Table \thetable.} Demographic characteristics of the 140 women who completed the recruitment survey, including age range, education level, employment status, and preferred communication language.\par}

\centering
\begin{tabularx}{\textwidth}{@{}lXrr@{}}
  \toprule
  \textbf{Survey Question} & \textbf{Response} & \textbf{N} & \textbf{Percent} \\
  \midrule
  \multicolumn{4}{@{}l}{\textbf{\textit{Age Range}}}\\
  \cmidrule(lr){1-4}
   & 18--24 & 101 & 72.1\% \\
   & 25--34 & 39  & 27.9\% \\

  \addlinespace[2pt]
  \multicolumn{4}{@{}l}{\textbf{\textit{Education}}}\\
  \cmidrule(lr){1-4}
   & High School         & 61 & 43.6\% \\
   & Bachelor's Student  & 16 & 11.4\% \\
   & Bachelor's Degree   & 58 & 41.4\% \\
   & Other               & 5  & 3.6\%  \\

  \addlinespace[2pt]
  \multicolumn{4}{@{}l}{\textbf{\textit{Employment Status}}}\\
  \cmidrule(lr){1-4}
   & Non-Employed         & 79 & 56.4\% \\
   & Fixed-Term or Intern & 28 & 20.0\% \\
   & Self-Employed        & 17 & 12.1\% \\
   & Other                & 16 & 11.4\% \\

  \addlinespace[2pt]
  \multicolumn{4}{@{}l}{\textbf{\textit{Preferred Communication Language}}}\\
  \cmidrule(lr){1-4}
   & English        & 48 & 34.3\% \\
   & Persian (Dari) & 84 & 60.0\% \\
   & Other          & 8  & 5.7\%  \\
  \bottomrule
\end{tabularx}
\end{minipage}
\end{center}

\section*{Appendix C: Employment, Online Learning, and GenAI}
\label{app:employment-online-learning-genai}

\begin{center}
\includegraphics[width=0.78\textwidth]{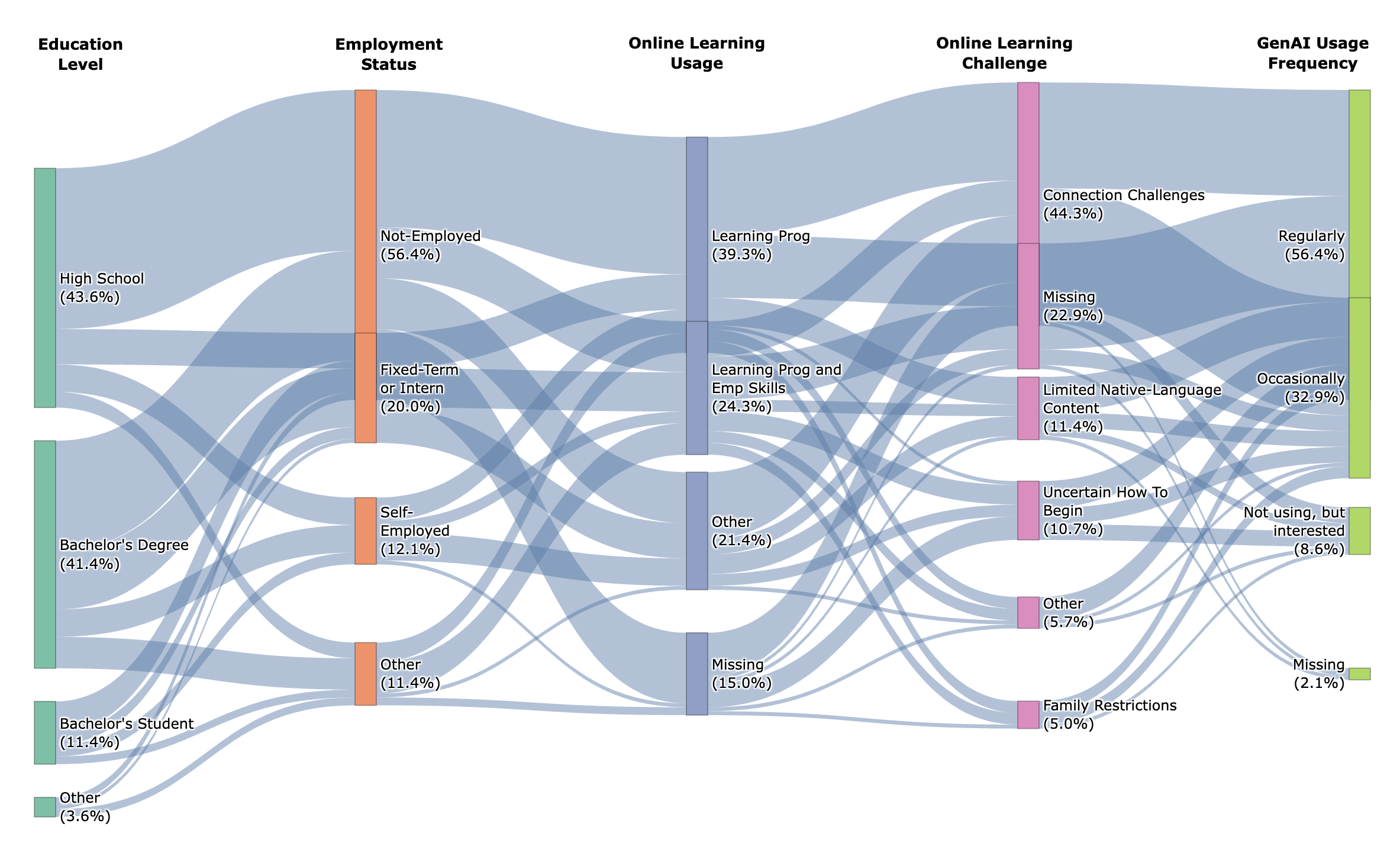}
\par
\begin{minipage}{0.84\textwidth}
\small
\refstepcounter{figure}\label{fig:DemoData}
\textbf{Figure \thefigure.} Demographic surveys overview showing flows between participants’ education level, employment status, online learning usage, key challenges with online learning, and frequency of GenAI use.
\end{minipage}
\end{center}

\section*{Appendix D: Storyboard}
\label{app:storyboard}

\begin{center}
\includegraphics[width=0.54\textwidth, angle=90]{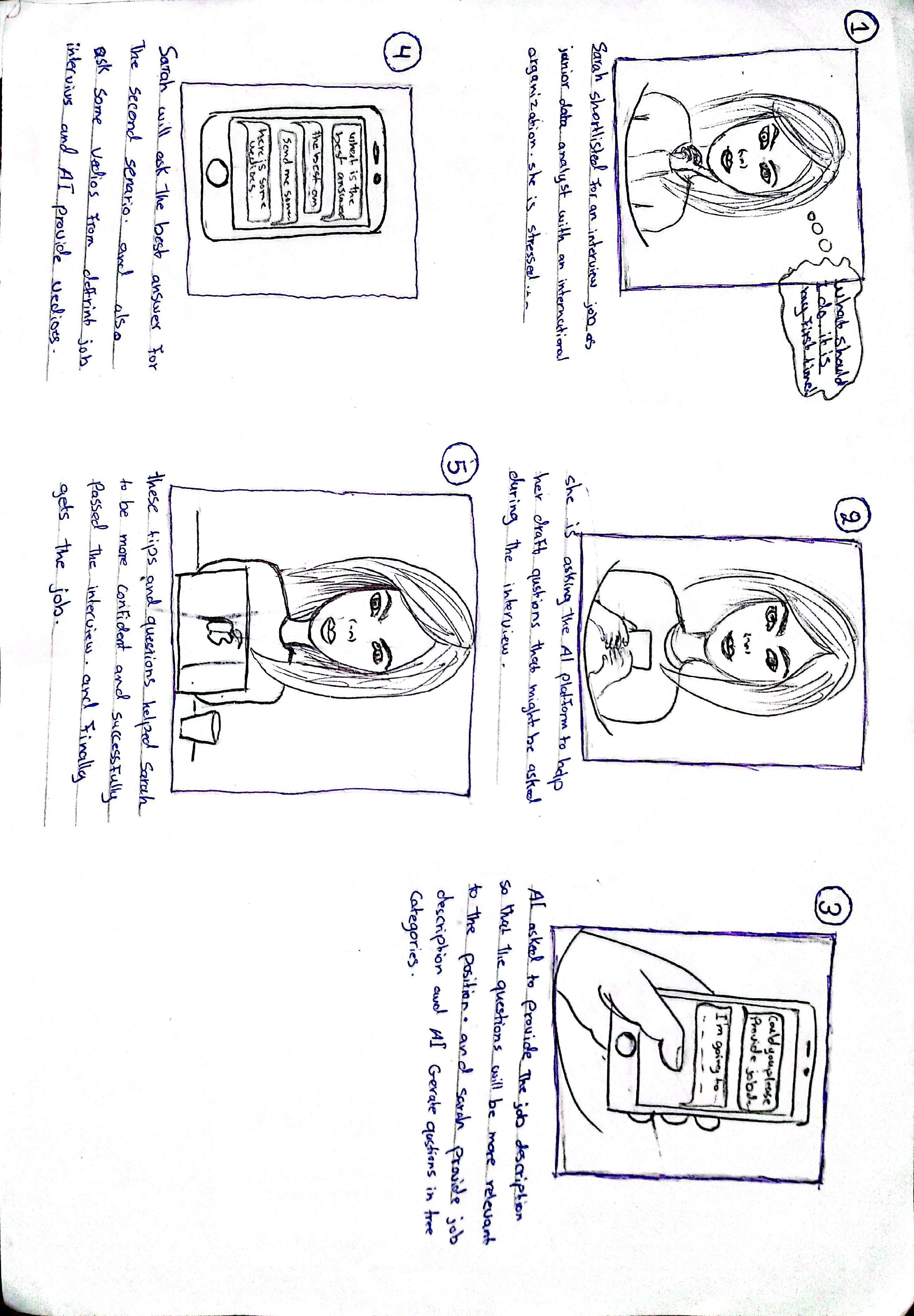}
\par
\begin{minipage}{0.76\textwidth}
\small
\refstepcounter{figure}\label{fig:Storyboardp}
\textbf{Figure \thefigure.} An example storyboard received from PD2\_4 based on the scenario she generated in open discussion.
\end{minipage}
\end{center}
% \twocolumn
\end{document}